\documentclass{article}

\usepackage{arxiv}

\usepackage[utf8]{inputenc} 
\usepackage[T1]{fontenc}    
\usepackage{hyperref}       
\usepackage{url}            
\usepackage{booktabs}       
\usepackage{amsfonts}       
\usepackage{nicefrac}       
\usepackage{microtype}      
\usepackage{lipsum}

\usepackage{cite}
\usepackage{amsmath,amssymb,amsfonts}
\usepackage[utf8]{inputenc}
\usepackage[T1]{fontenc}
\usepackage{multirow}
\usepackage{hhline}
\usepackage{float}
\usepackage{graphicx} 
\usepackage{caption} 
\usepackage{subcaption} 
\usepackage{siunitx}
\usepackage{footmisc}

\title{A Machine Learning Approach to Predicting Continuous Tie Strengths}

\author{
  James Flamino$^{\dagger}$ \\
  Department of Physics\\
  Rensselaer Polytechnic Institute\\
  Troy, NY 12180 \\
  \texttt{flamij@rpi.edu} \\
   \And
  Ross DeVito$^{\dagger}$ \\
  Department of Computer Science and Engineering\\
  University of California San Diego\\
  La Jolla, CA 92093 \\
  \texttt{rdevito97@gmail.com} \\
  \And
  Boleslaw K. Szymanski \\
  Department of Computer Science\\
  Rensselaer Polytechnic Institute\\
  Troy, NY 12180\\
  \texttt{szymab@rpi.edu} \\
  \And
  Omar Lizardo \\
  Department of Sociology\\
  University of California Los Angeles\\
  Los Angeles, CA 90095 \\
  \texttt{olizardo@soc.ucla.edu} \\
}

\begin{document}
\maketitle

\begin{abstract}
Relationships between people constantly evolve, altering interpersonal behavior and defining social groups. Relationships between nodes in social networks can be represented by a tie strength, often empirically assessed using surveys. While this is effective for taking static snapshots of relationships, such methods are difficult to scale to dynamic networks. In this paper, we propose a system that allows for the continuous approximation of relationships as they evolve over time. We evaluate this system using the NetSense study, which provides comprehensive communication records of students at the University of Notre Dame over the course of four years. These records are complemented by semesterly ego network surveys, which provide discrete samples over time of each participant's true social tie strength with others. We develop a pair of powerful machine learning models (complemented by a suite of baselines extracted from past works) that learn from these surveys to interpret the communications records as signals. These signals represent dynamic tie strengths, accurately recording the evolution of relationships between the individuals in our social networks. With these evolving tie values, we are able to make several empirically derived observations which we compare to past works.
\end{abstract}

\footnotetext[2]{Authors contributed to this work equally}

\section*{Introduction}

Relationships and the interactions that characterize them are a defining features of social networks~\cite{borgatti2009network,kitts2020rethinking,rivera2010dynamics}. In the network and social sciences, the strength of these relationships are often represented by a ``tie strength,'' a weighted edge between two nodes that marks the existence of a connection between the people portrayed by the nodes. Previously, work on understanding tie strength has ranged from interpreting its importance in information spread~\cite{granovetter1977strength,krackhardt2003strength,marsden1984measuring,policarpo2015friend,onnela2007structure} to using a variety of social features to predict the magnitude of tie strength between individuals~\cite{jones2013inferring,gilbert2009predicting,conti2011a_model,wiese2015you,mattie2018understanding}. 

A question fundamental to this topic is: what contributes to the strength of a tie between two people? Or, more specifically, what attributes of a relationship can we use to predict a tie strength value that properly represents the closeness of two individuals within a social network? This question has no singular answer, though there have been popular works delving into possible interpretations~\cite{granovetter1977strength,krackhardt2003strength,policarpo2015friend}. Such works have pointed to both qualitative and quantitative attributes of relationships that seem to influence the strength of a the relationship between two individuals, and therefore would contribute to the evolution of tie weights within the involved social network. 

In Granovetter's popular early work on this topic~\cite{granovetter1977strength} these factors were identified as time invested, emotional intensity, mutual confiding, and reciprocal services. He suggested tie strength was ultimately a linear combination of these factors. Krackhart's response to this work~\cite{krackhardt2003strength} introduced an alternative characterization of tie strength that consisted of interaction frequency, affection, and time, which he defined qualitatively as an enduring history between the two linked individuals. 

Marsden's work provided further clarification to the considered factors, introducing predictors (aspects of relationships that are related to, but not a part of, tie strength), and indicators (actual components of tie strength). The former set of factors contain relationship descriptors like kinship and educational differences. The latter set of factors contained attributes of communication and shared interests, and intimacy that are more commonly seen as features of tie strength in other works. In particular, Marsden addressed closeness (emotional intensity), duration of connection, frequency of communication, breadth of discussion topics, and mutual confiding (all of which correspond to Granovetter's characterizations of tie strength) and found that closeness played an important part in informing tie strength.

Given the subjective nature of relationships and their attributes like intimacy and affection, a more robust and generalizable quantitative approach to prediction faces some challenges, though there has been groundwork laid to this end~\cite{gilbert2009predicting,jones2013inferring,conti2011a_model,wiese2015you,mattie2018understanding}. In some of these works, tie strength is approximated by linking features like communication frequency, social media friend overlap, shared attributes (like gender or education), directed message keyword usage, and the like to predict closeness between two individuals. The predictions are then usually compared to a ground truth extracted from a survey asking participants to rate their closeness either on a numbered scale or indirectly by using questions like ``How strong is your relationship with this person?''.

Facebook has become a prevalent medium for these kinds of experiments. Since among the online social media platforms, Facebook maintains a massive social network and facilitates broad forms of interaction between users. The results of these experiments revealed that features like days since the last communication, participant's number of friends, and exchanged intimacy words contribute a fair amount to the prediction of tie strength. In addition to this, some of the work showed that public communications, like Facebook wall posts, and private communications, like private Facebook direct messages, often contribute equally to predict tie strength. 

Despite the interesting implications of these works, the scope of these systems are always limited, restricted to a snapshot in time of the social network. But as most people's lives are constantly a witness to, relationships evolve over time. They can be subject to changes, and such changes will have a direct impact on how people interact with former, current, and future friends. In fact, the progression of a relationship is important for characterizing the connection's strength. All of the past works mentioned above only focus on predicting tie strength at a single moment in time. Additionally, these models often faced the issue of being tied to their specific application. The representations of tie strength were often characterized by attributes extracted from a singular platform, namely Facebook. This results in interpretations of tie strength that are defined by their specific platforms, making them incapable of being applied generally.

In this paper, we lay out a generalizable system that addresses these concerns and demonstrates that evolving tie strengths for a dynamic social network can be accurately predicted given only a practically small survey-based ground truth. There are three core pieces to our system: the input data, the training data, and the model that learns to interpret the input data as tie strengths using the training data. 

A person's digital communication records are used as the input data, from which a trained model can predict their social ties. While communication is just one of the many hypothesized aspects of a relationship which impacts the tie strength, as a data source, digital communications have the practical benefits of being abundant, multifaceted, and easy to collect. These advantages grow as the world becomes increasingly dependent on digital interactions. Our system can work with any number of communications mediums simultaneously (e.g. text messages, phone calls, video calls, WhatsApp, and Facebook Messenger), it just requires that the records include the time, type, and pair of people involved for each communication event.

To train a model and evaluate its performance on converting input data into tie strength values, ground truth data on social ties is needed for some subset of those for whom we also have communication records. To meet this need, our system just requires a small number of top $k$ lists of social ties. These lists can be procured at any time over the course of the dataset, as long as there is concurrent communication data related to the person who's top social ties make up the list. Any representation of tie strength ground truth in practice would likely be from survey responses. This being the case, a ranking-based ground truth would be more robust when compared to more common exact social tie values or binary relationship labels. Using this ordering benefits from never having to ask for explicit social tie values or cutoffs, which is important as these concepts would be highly subjective among survey respondees. Furthermore, we show that even asking for an explicit ranking in the survey is not required to avoid response bias. Instead we derive our top $k$ rankings using survey questions based on Granovetter's and Krackhart's work.

We found pairwise comparison based machine learning models to be excellent predictors in comparison to the baselines. These models are able to take advantage of using top $k$ orderings and pairwise comparisons to provide many training examples for their underlying models using a realistic amount of survey based ground truth data. This is important as machine learning performance tends to rise with an increase in quality training data. This pairwise comparison framework has the additional benefit of producing an interpretable social tie value.

We train these machine learning models using the top $k$ lists, and show that once trained, we can use these models to continuously, and accurately, predict the evolving tie strength of one person towards another using just communication data. In the following sections we discuss our system in greater detail, evaluating its efficacy at accurately producing evolving tie strength values. We then show that analyzing these dynamic values for all participants reveals interesting observations related to communication evolution, relationship stability, and triadic dynamics.

\section*{Data}

To develop and evaluate our dynamic tie strength model, we needed data on a social network, and data on relationship attributes that could be used to predict tie strengths within that social network. Optimally, both would extend over a long enough period of time to capture changing relationship dynamics. The NetSense~\cite{purta2016experiences} study, which consist of data voluntarily collected from randomly selected students entering the the University of Notre Dame, fits this need, providing linked ego network surveys and digital communications records. Data for this study was collected from Fall 2011 and Spring 2013. Student phone records, including text messages and phone calls, were provided, along with corresponding ego network surveys that were filed out each semester by the participants. In terms of scope, NetSense followed 196 students at its peak, yielding extensive communication records that we could use to capture relationship changes over time.

\subsection*{Communication Records} 

NetSense's communication record conforms to the standard Call Detail Record (CDR) format, listing a timestamp, sender, receiver, message type, and message length. The NetSense study contained $7,465,776$ events generated by the participants of the study. Text messages make up about $94\%$ of these events, with the remainder being calls. Despite this imbalance in volume, phone calls remain an important medium for communication and carry an emotional weight, especially among the younger population captured in the study~\cite{blair2015cell}. Thus, we choose to include both calls and text messages. In fact, we find that considering calls and text separately also improve the machine learning models we implement.

\subsection*{Ego Network Surveys} 

As mentioned earlier, ego network surveys were collected once a semester to complement the communication record. These surveys were prefaced with a question asking the survey-taker (the ego) to list individuals (the alters) with whom they spend a significant amount of time communicating or interacting. This list could include up to 20 alters, and could include people that were not involved with the study. This allowed for these lists to contain a variety of relationship types, including fellow students, roommates, parents, siblings, coworkers, and romantic partners. The ego was subsequently asked to specify their relationships with these individuals. This classification was provided to the ego as a closed list. In general, options available ranged in familiarity from ``significant other'' and ``parent'' to ``acquaintance''. Other related information on these alters was also collected through additional follow-up questions that included asking about the history of contact, shared interests and activities, and the frequency of communication. Importantly, the surveys also asked the ego to subjectively rate similarity and closeness with the alters.

Despite the thoroughness of this study, the time between survey postings is significant: the data has four ego network surveys over the four semesters, and the study participants listed on average $15.9$ people per survey. Regardless of the sparsity of these ego network surveys, our results demonstrate that they still provide sufficient ground truth support for our models.

\subsection*{Definition of Tie Strength}

Using Granovetter's and Krackhart's tie strength definitions, we can outline a template for evaluating the connection between an ego and their alters. In early works on this subject, tie strength was represented discretely as labels (such as ``close'' or ``not close''). More recently, tie strength has been encoded in the form of a numerical range, which conforms to Granovetter's belief that tie strength is continuous, not discrete. These representations were varied, and were often based on a combination of some large set of qualitative and quantitative predictors. However, these methods for interpreting tie strength are also usually reliant on platform-specific attributes~\cite{gilbert2009predicting,jones2013inferring}. 

To avoid these limitations, we take a different approach to encoding tie strength by simply ordering the list of alters from our ego network surveys. For any given ego network survey, we produce a ranked list of the individuals listed in the survey, where the order is determined by how strong the survey-taker's social tie is with each individual. Subsequently, each survey-taker produces a set of top $k$ social tie rankings, timestamped by their respective ego network surveys (greater details of how this is done are presented in the methods section). We train and evaluate our social tie prediction models through how well their predicted tie strengths conform with these rankings at their corresponding times. Specifically, we introduce a suite of models that interpret our communication data streams as dyadic tie strengths. These tie strengths are represented by a signal value, which is used to establish a predicted ranking by ordering said signals by magnitude. We compare this ordering with the corresponding survey's ground truth top $k$ social tie ranking.

Given that these rankings are determined by how close an alter is to their ego, a model that is properly trained to produce signals that accurately reconstruct the appropriate ranking of each alter for any associated ego ultimately means that the model is capable of generating continuous signals that are a representation of evolving tie strength between an ego and those they've communicated with. In other words, a signal's magnitude that indicates the level of affinity an individual has for another in the context of a ranked list fits within the definition of evolving tie strength, which in the past has been defined loosely. And since the model is designed to generate signals over long periods of time (as tuned by the multiple social tie rankings over time in the training data), any new target individual with simple communication data should be able to have the model produce tie strengths that (when ordered by magnitude) are be able to effectively identify their closer social ties and subsequently capture the evolution of their connections with the new target individual across the course of their communication data.

\section*{Models}

Our suite of models for this survey reconstruction process can be divided into two classes: a baseline class and a machine learning class. For the baseline class, we implement single-attribute models that use specific attributes of communication behavior that are often cited as aspects of tie strength or directly used as a proxy for it~\cite{conti2011a_model, wiese2015you, gilbert2009predicting, granovetter1977strength, mattie2018understanding}. The second class, the focus of this paper, uses machine learning methods on time series or a collection of single-attribute model values to predict all out tie strengths for a target person whose communications record are given. These models do this by making pairwise comparisons between everyone the target person had communicated with.

\subsection*{Baseline Models}
It has long been postulated ``the more frequently persons interact with one another, the stronger their sentiments of friendship for one another are apt to be''~\cite{homans_1950}. Previous research has often used the frequency of communication to predict tie strength or emotional closeness~\cite{conti2011a_model, wiese2015you, gilbert2009predicting}. Following this established methodology, we created a frequency model that calculates frequency by dividing the number of communication events between two individuals by the elapsed time since they first communicated at the timestep for which it is being evaluated. In addition, we assessed a recency model that uses the elapsed time since the last communication between two people as an inversely related estimate of frequency of contact, as is done in~\cite{conti2011a_model}. This measure of time since last communication was found to be the most predictive single feature in~\cite{gilbert2009predicting}. 

Stronger ties, by definition, tend to involve longer time commitments~\cite{granovetter1977strength}. Following this logic, we also created a duration model that uses the time since the first communication record as a proxy for length of friendship or other social bond. This was found to be the second most predictive feature in~\cite{gilbert2009predicting}. As another proxy for a pair's time commitment to communicating, we add the volume model, which counts the total number of calls and text messages between two individuals.

Recently, there has been work showing that tie strength can also be predicted using the overlap of friend groups between two specific individuals~\cite{bott2014family}. In one implementation of this concept~\cite{mattie2018understanding}, a metric called ``weighted overlap'' for social bow tie structures is used as a feature to help machine learning algorithms predict the tie strength between two individuals in a specific time frame. Given that this particular feature contributes heavily to the predictive performance in a couple of the tested cases in this work, we implement this metric here as well as a model to explore the predictive capabilities of evolving friend group overlap. The specific implementation is shown in detail in the methods section.

Given that this is the baseline class, we also ensure to set the lowest bar for survey reconstruction with a simple random baseline. This randomly sorts the individuals with whom the target participant had communicated previously into an arbitrary ranking.



\subsection*{Machine Learning Models}

At their core, our machine learning models compare a selected individual's (person A) communication history with one individual (person B) against A's communication history with another (person C). Provided these two histories, the machine learning models then will predict, between B and C, which of the two will have a greater tie strength with A. When these comparisons are made for all pairs of people in the selected individual's records, we can generate the predicted ranked list for that person (see Methods for more details), and subsequently produce meaningful tie strength values for all of this person's relationships. 

These tie strengths for the machine learning models are expressed as \textit{winning percentages}. For an individual being evaluated, winning percentage is the fraction of pairwise comparisons with all other people the target had communicated with were the model predicts the evaluated individual has a stronger social tie. This score has the range $[0,1]$ where higher the score means the more likely the scored individual is closer to the selected individual. This tie strength value can also be generated at any point in time provided the models are trained and there is communication history available for those being considered in the pairwise comparisons. 


Pairwise comparison-based ranking models can also take advantage of a ground truth in ranked form. Specifically, selecting permutations from this ordering allows for the generation of many training examples from relatively little surveying. This is important as the quality of machine learning models is tied to the quantity and quality of training samples.

The first machine learning model uses an ensemble method that utilizes features of duration, recency, frequency, and volume in the communication data to inform a random forest classifier~\cite{breiman2001random}. The random forest classifier predicts which of the two compared connections with a selected individual is indicative of a greater tie strength, and thus should be higher ranked. The second machine learning model uses communications time series and recurrent neural networks, specifically a two-channel Long Short Term Memory (LSTM) networks~\cite{hochreiter1997long}. The LSTM is used to make pairwise comparisons, which are in turn used to produce a signal and ranking, all of which is done in the same manner as in the Ensemble model. As mentioned earlier, we find that performance improves for both machine learning models when texts and calls are treated as separate data streams.

\section*{Results}

\subsection*{Ranking Metrics}

To determine survey reconstruction error and evaluate the models' capabilities, we compare a model's predicted ranked individuals against the ground truth using the rank-biased overlap (RBO)~\cite{webber2010similarity}. RBO is an indefinite rank similarity measure used to evaluate the similarity of two ranked incomplete lists, making it better suited for this task than, for example, Jaccard. RBO has several other desirable attributes for this kind of comparison; it handles items being present only in one ranking, weights higher ranking items more, works with any given ranking length, and requires little to no assumptions about the data. 

\subsection*{Ranking Performance}

The evaluation process we chose for comparing our suite of models against the NetSense ground truth was the standard 3-fold cross validation. Given the total list of NetSense egos, the validation process shuffled and equally split this list of participants into three mutually exclusive groups. For each fold, a test group was selected, with the remaining two groups used as training data. Within a fold, we separate the training and testing data into four subsets, split equally by semesterly survey time (the time at which the surveys were filled out by the egos). At each survey time, the training and testing subset only includes the surveys and communication data from before that time (therefore the later subsets contain the earlier subsets). Then for each subset we trained and tested the machine learning models with all training data available. We then subsequently tested these trained models on the available testing data in the subset, using the models to predict the current testing ego's surveys. This process was repeated for each subset, allowing more training and testing data to be released with each proceeding survey time. We do not, however, allow the models to train on the preceding ground truth of the test data after it's been predicted. Instead, we ensured the models have predicted the surveys for the test egos for each survey time first within the current fold before releasing ground truth for comparison. This setup prevents any model from using future communication history in any of the folds during evaluation. Once this setup was completed for each model within the current fold, the fold score was determined by finding the weighted average survey reconstruction accuracy (RBO) across all available surveys for each individual test participant. The weight of each predicted survey is the size of that survey's ground truth, accentuating the prediction of larger surveys. Given a score for every participant in this fold, this score was then averaged over all test participants. The final score was computed as the average fold score over all folds, which are shown for all models in Table~\ref{tab:performance}. 

Of the baseline class models, overall volume of calls and texts since the start of college was the most predictive. This was followed by frequency and recency of communication. Duration of communication was a relatively distant fourth, but this may be tied to the time frame of data to which we had access to. Specifically, those with whom a study participant had been friends with long before college could only have an estimated duration of friendship spanning back to the start of college. This start of college period was also a time when participants were making many new contacts. Some of these contacts would go on to become friendships, but many were just freshmen meeting new people who would not be significant as their time in college went on. For this reason, this may have been a poor estimate of the length of friendship and therefore of the social bond. 

The limited scope of the data negatively affected the overlap model as well, which performed even worse than the duration model. The limit in accuracy here is most likely from the fact that the communication data only provides the comprehensive communities of neighboring friends for study participants. Non-participants do not have their out-going messages recorded, so their only neighbors will always be strictly participants. The overlap model requires a sizable sample of the overlapping and non-overlapping neighbors of both individuals being evaluated for tie strength prediction. As one of these two individuals might be a non-participant, their neighborhood will be incomplete which skews the overlap value. These issues demonstrate some of the difficulty of inferring social ties given just a single target person's records.


\begin{table}
\centering
\caption{Results of the NetSense survey reconstruction models, with variance in parentheses}
\label{tab:performance}
\begin{tabular}{|cl|l|} 
\hline
\textbf{Model Class} & \multicolumn{1}{c|}{\textbf{Model}} & \multicolumn{1}{c|}{\textbf{RBO}} \\ 
\hline
\multirow{6}{*}{Baseline} & Random & 0.037 (0.003) \\
 & Overlap & 0.064 (0.008) \\
 & Duration & 0.234 (0.033) \\
 & Recency & 0.307 (0.025) \\
 & Frequency & 0.320 (0.032) \\
 & Volume & 0.363 (0.032) \\ 
\hline
\multirow{2}{*}{Machine Learning} & Ensemble & 0.450 (0.029) \\
 & LSTM & 0.481 (0.029) \\
\hline
\end{tabular}
\end{table}

But despite these limitations, the Ensemble and LSTM models produced RBO scores of $0.450$ and $0.481$ respectively for the NetSense study. While frequency, recency, duration, and volume of communication have merit for approximating relationship strength on their own~\cite{homans_1950, conti2011a_model, gilbert2009predicting}, they are unable to capture a greater whole of the latent mechanics of social dynamics. But when all used in conjunction, this feature space allows for even a simple random forest classifier to perform well. However, the relative performance of the Ensemble model hints at the weaknesses of such simple models for inferring complex social dynamics, and that even more improvements can be made. This is where the final model, the LSTM comes in.

Recurrent neural networks can accept, as input, temporal sequences of an arbitrary length. This allows them to use as features the whole communications histories from the start of the dataset to any time at which social ties are being evaluated. This ability to analyze histories of interactions in a temporally aware way is likely the key to achieving the best performance. Instead of using heuristic features calculated at a specific time, the LSTM is able to internally learn latent features from patterns of communication over time that are most meaningful for evaluating the strength of social ties. 




\section*{Evolving Tie Strength Analysis}

\subsection*{Evaluating Continuous Signals}

\begin{figure*}[tb!] 
    \centering
    \makebox[\textwidth]{\includegraphics[width=.8\paperwidth]{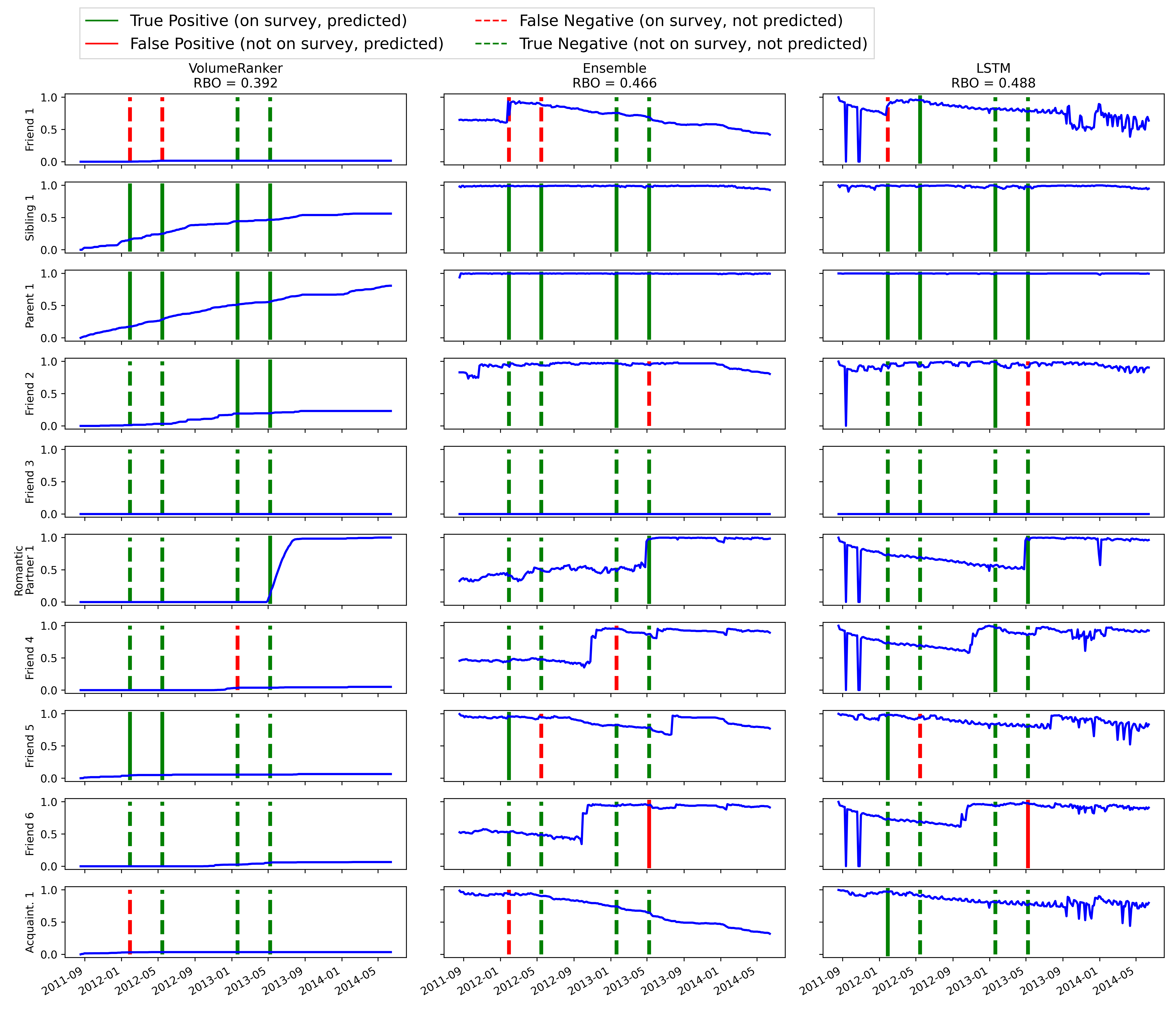}}
    \caption{Generated signals for a randomly selected NetSense participant using the top three survey reconstruction models (Volume, Ensemble, and LSTM). The x-axis marks the timestamp, and the y-axis marks the signal magnitude. The colored vertical bars indicate the occurrence of a survey, denoting if the models correctly (or incorrectly) classified the ranking of the considered individual.}
    \label{fig:signals}
\end{figure*}

As mentioned in the Data section, the signals that can be generated by the trained models are used to represent evolving tie strengths due to the fact that the ground truth rankings that the models are fit to are ordered by closeness. Hence, the signal magnitude between two individuals is also the magnitude of tie strength between them. Now, to illustrate the signal generation dynamics of our top performing models, we present the evolution of tie strengths between a NetSense study participant and a sample of their listed alters as encoded by our Volume model, the Ensemble model, and the LSTM. For consistency, we normalize all values. In this particular analysis we train the machine learning models using all the NetSense data, excluding all data relating to the selected participant. We sequentially sampled the resultant signal values of our two trained machine learning models and the Volume model over the entire duration of the NetSense study, which includes the time of each survey where the ground truth values are known. We plot these sampled values against time, marking the times of the surveys and denoting if the signals, when ordered with all other alters in ascending order by value, place each alter at the right survey rank when compared against ground truth. These signals, as generated by the top four models for a selected participant, are shown in Figure~\ref{fig:signals}. We specifically selected a subset of listed individuals that had particular relationships with the target individual (e.g. parents, siblings, significant others, and close friends). 

The first observation that we must be made in Figure~\ref{fig:signals} is the differences in signal shapes between model types. Naturally this is due to the differing signal generation methods of each model. Specifically, the Volume model captures the continuously growing intensity of  communication while the machine learning models convert their pairwise predictions into signals using comparative probability (i.e. a strong signal for an individual means they have a higher probability of being closer to the selected participant than other individuals).

But despite the differences in tie strength interpretations between models, there are still clear trends that are reflected across models. For example, in Figure~\ref{fig:signals}, ``Romantic Partner 1'' becomes socially involved with our selected participant right around the time of the last survey. This is universally reflected by a massive spike in predicted tie strength, which is maintained for the remainder of the study. The consistent communication between the selected participant and their family (the sibling and parent) is also easily shown across the models, with accompanying high tie strength values.

The universal visibility of these relationship transitions can be attributed to an accompanying spike in communication activity. Since the Volume model operates directly using total event occurrences, obvious shifts in communication behavior are easily captured in the outputted signals. However, only the machine learning models are able to capture more nuanced trends visibly. 

One example can be found in the listed individual ``Friend 4''. Around the 3rd semester, this friend and participant became strong friends, enough to warrant the participant placing the friend on their ego network survey. But communication is sparse overall as shown by the Volume model, meaning a change in calls and text volume was not the biggest change here. An underlying shift in the pattern of communication occurred in a way that only the machine learning models were able to detect it, and subsequently boost the tie strength value between the two. We can use ``Acquaintance 1'' as another example. This listed individual initially meets the participant around the first semester, most likely through a study group. Beyond the first semester the participant forms other more concrete and significant social circles, relegating the acquaintance to a strictly academic role (which is why communication with them persists past the first semester though they are no longer included within the surveys). This social tie weakens as class overlap inevitably diverges, and the two eventually move on with their lives. While these initial anecdotes are interesting, analyzing the tie strength dynamics of larger groups could allow us to draw stronger general conclusions. 

For the remainder of our analysis, we will be using the LSTM model with its winning percentage representation of tie strength. In addition to being the best performing models in our suite, its comparative winning percentage representation of dynamic tie strength is meaningful and easy to understand. For this analysis, we sampled predicted tie strengths from the LSTM model across the duration of the NetSense study. We can represent the inferred strength of a social tie from one person (who's communications records are being used for the inference) to another as a directed edge between nodes. The weight for each edge is the tie strength value itself, which varies as a function of time. With the resulting dynamic social network, we can analyze general tie strength trends and compare them with previous work to verify the efficacy of the model in capturing important relationships trends using our generalizable methodology of survey reconstruction.

To establish the groundwork for future research, we deliberately chose two broad subjects for initial analysis: dynamics of relationships and dynamics of triadic groups. The analysis of tie strength evolution as it relates to different relationship types is important, as the different kinds of classifications (i.e. friend, sibling, parent, etc) can often help determine the trends of closeness between individuals in the past and future. We also choose to analyze triadic motifs within our evolving network due to the importance of triads in distinguishing communities~\cite{seshadhri2012community} and network structure as a whole. 

\subsection*{Relationship Dynamics}

As stated in the Data section, participants were asked to classify their relationships with those they listed in their ego network surveys. Given these classifications, we can analyze how signals on average change with time for different nodes depending on their relationship type. In particular, we analyze the average edge weight over time for friends, kin, and significant others (as identified by survey-takers), to evaluate relationship stability over time. 

\begin{figure*}[tb!]
    \centering
    \includegraphics[width=\textwidth]{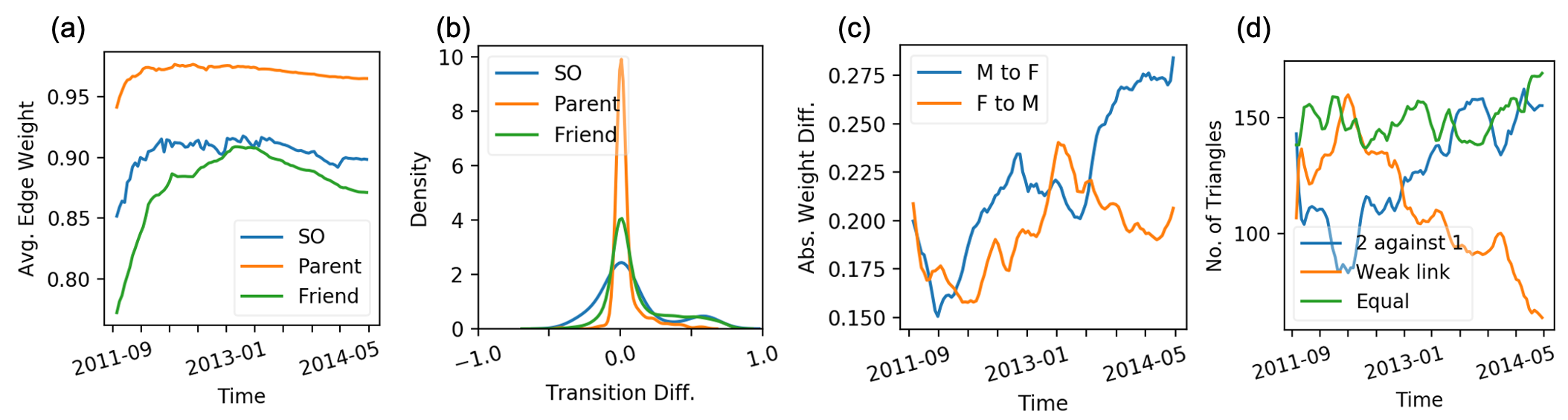}
    \caption{A sequence of illustrations for the analysis of the evolving tie values in NetSense as generated by one of our best survey reconstruction models. (\textbf{a}) shows the growth of the average edge weight of different relationships over time. (\textbf{b}) shows the Kernel Density Estimation of transition differences between relationships. (\textbf{c}) shows the absolute difference in edge weight between gendered majorities and minorities in detected triads in our social networks. (\textbf{d}) shows the number of detected triads that fit into one of the three standard triadic motifs.}
    \label{fig:network_analysis}
\end{figure*}

In Figure~\ref{fig:network_analysis}a we show the average edge weight over time for our selected relationship types. We found that while friends tend to have a more volatile edge weight (due to the fact that there are so many different kinds of friends), parents (and similar kin) tend to stay fairly consistent with high tie strength. This agrees with previous research~\cite{chen2009extending}, which shows that college students very often maintain consistent contact with their families. For those that have frequent communication with their parents and siblings, this behavior is correlated with their closeness to them, as reflected in survey rankings. 

The average tie strength value is also very stable towards parents across time, with little variation (which can be even be seen in Figure~\ref{fig:signals}), indicating that the rank held by parents rarely wavers between participants, confirming the general strength of connection. This is likely due to the fact that if parents are even going to make it on to a ranked list of individuals with whom the participant has communicated with the most, these relationships are going to inherently stable. If they were not, students would not invest time communicating with them enough to warrant listing them. 

Surprisingly, significant others are often times ranked below parents. This difference in ranking is due to the fact that the label of romantic partner is not static, like the label of ``parent.'' Significant others can be introduced prior to the study or anywhere during the study, and that label can be removed or changed at any time as well. This effect is seen to a greater degree with the term friend. In addition to the conditional assignments of the term, there are many different stages of rapport that any friend could be while being included in a participant's survey listing. It is also important to note that despite the volatility (or lack thereof) of these classifications, the growth of tie strengths inevitably slows as time goes on. This is reflected in Figure~\ref{fig:signals}, as most signals settle after either a transition point or a period of growth/loss. This indicates in the absence of a perturbing force (e.g. getting to know a person, experiencing a breakup), relationships tend to cement themselves to some standard level closeness, resulting in some consistent pattern of communication. A similar observation was made in Saramaki's work~\cite{saramaki2014persistence}. In this paper, social signatures between individuals in a social network were derived directly from communication data. These social signatures were found to be generally stable and consistent in shape, corroborating the trend of stability seen in tie strength signals here.

We can further characterize our analysis of relationship interactions by analyzing the transition difference in signal weight over time. We do this by finding the greatest change in signal value (the ``transition point''), then taking the difference in the average signal before and after this transition point. We find the transition difference for every listed individual of the study participants, with the resultant values binned by relationship label. The Kernel Density Estimation (Gaussian kernel) of the binned transition differences for Significant Others, Parents, and Friends is seen in Figure~\ref{fig:network_analysis}b. 

We find that for our data, most family-related relationships remain stable with a primary mode centered about $0$, and a very slight mode about $0.5$, indicating there is usually either little change or deviation from whatever the initial signal was, or if there was change, it was positive. Significant others in NetSense had a more noticeable positive trend, with a primary mode at $0.5$ and a secondary mode at $0$, with a tail that trails off into the negative region. This shape marks the dynamic of close (but non-familial) relationships, which undergo positive changes when the relationship forms, and negative changes when a break-up occurs. And since this kind of relationship is a lot more volatile than a family-related relationship, this type of dynamic occurs more often, and is therefore reflected more by our models. The distributions of transition differences for friendships, like significant others, has a wider mode at $0$, with a smaller mode at $0.5$ and a tail into the negative domain. This shape can be attributed to the varying types of friendships that can be classified under the ``friends'' relationship class in the ego network surveys. An example the variety of friends and their associated signals for this can easily be seen in Figure~\ref{fig:signals} as well, where the signal shapes for the friend classifications differ visibly, as opposed to the signals of the kin classifications. Importantly, these significant stability differences between friendships and kin relationships have been observed before~\cite{roberts2011costs}. In this work, kin relations were found to be more stable and maintain a higher level of emotional closeness with little maintenance, compared to friendships which were less stable and required active maintenance to prevent decay. Our analysis further confirms these observations with the clear quantification of the difference in tie strength stability in friends versus family.

\subsection*{Triadic Dynamics} To analyze the dynamics of triadic motifs within our network, we first extract a set of triads that consistently occur in the communication data across each semester. With these mappings we then analyzed the tie strengths for the in-degree and out-degree edges between each of the nodes involved. With the mappings and associated the tie values obtained, we choose to first focus on the evolution differences between genders within triadic groups. In particular, we looked into mixed gender triads, analyzing groups where there were two males and one female, or two females and one male. We took the absolute difference between the average degree value of the majority gender and the average degree value of the edges from the majority gender to the minority gender. This difference characterizes how the majority treats the minority and evaluates the interconnected relationships. The differences across times for both triad types is presented in Figure~\ref{fig:network_analysis}c. As seen in the plot, majorities do often interact with minorities differently, though male majorities do so to a greater degree than female majorities, though not by a great amount. The interactivity difference between a triadic majority and minority can be viewed as two nodes uniting against one (either directly or indirectly). This two-against-one behavior is fairly common in sociology~\cite{caplow1968two}, and happens to varying degrees. But how prevalent is this motif?


To answer this question, we can use our evolving tie strength values to observe the growth of this motif, and compare it to its counterparts: the weak link triad and the equalist triad. In the weak link scenario, two of the three nodes are highly connected; however, there is one link (in both directions) in the triad that is weak compared to the others. In terms of social group formation, this would mean that while two members are good friends with the third member (and vice versa), the two members themselves are not as strong friends with each other. Alternatively, in the equalist scenario, all links in the triad are fairly equal in value. At every timestep in our dynamic network, we tally the number of triads that meet one of the three criteria to track the trends of the triadic dynamics. These trends are shown in Figure~\ref{fig:network_analysis}d. We find that while often times equalist triads is the most prevalent, there is a consistently growing trend of two-against-one triads that become more established as time evolves. This is reflected in Figure~\ref{fig:network_analysis}c, as the absolute difference in edge weight between majority and minority grows. Overall, this indicates that the two-against-one motif is a prevalent dynamic in triads as tie strengths settle. Essentially, in a triadic dynamic, after friendships begin to cement, there will often be a ``third wheel''.

\section*{Discussion}

Tie strengths play an important role in the analysis of social networks, characterizing the relationship between individuals and providing insight into how those involved will interact with each other~\cite{granovetter1977strength,krackhardt2003strength,marsden1984measuring}. While past works have delved into predicting tie strength~\cite{jones2013inferring, gilbert2009predicting}, there has been limited research into the forecasting and subsequent analysis of tie strength that evolve over long periods of time. The paucity of this kind of research is in part due to the difficulty in collecting data social ties as they change, which is typically done through surveys. Additionally, many past works tend to implement tie strength measures that depend on many platform-specific attributes. In this paper we address both problems by introducing a system that converts easily collected communication data into continuous tie strength values with machine learning. We design this system to be generalizable, depending only on the communication data and a sparse number of ego network surveys. Using a small set of modular questions, we extract social tie rankings from the surveys that we use to train our predictive models and predict the social tie rankings. The trained models can also convert communication data to continuous signals over time. Given the nature of these signals that are generated by our models to predict survey rankings, we can interpret these values as continually evolving tie strengths. And with these values, we can analyze the relationship dynamics of social networks.

The NetSense study provided long term real-world communication data and surveys that provided ground truth values at points in time. Using machine learning, we are able to reconstruct ranked versions of these surveys with a relatively high average RBO. Provided the resultant continuous tie strength values from the best-performing models, we are able to effectively track the evolution of relationships (like identifying the time at which a significant other enters a participant's social circle). Furthermore, we show that relationships with parents (and other close family, like siblings) remain fairly consistent over time, with significant others coming in second in terms of tie strength stability. By further analyzing tie strengths about signal transition points we show that while parent tie strengths tend to be very strong and stable, often going unchanged over time, while significant others are more likely to experience significant transitions (driven by the initial formation of the relationship, or a subsequent dissolution). We also establish the paradigm that without a perturbing force, most relationships reach some form of resting-state as tie strengths settle. 

We further our analysis by looking into triadic dynamics of participants. We find that in mixed triads, male majorities tended to treat female minorities differently, while female majorities did the same to male minorities to a slightly lesser degree. This behavior reflects the two-against-one triad motif. We delve into this observation further by comparing this behavior against two other triad motifs (weak link and equality), and observe the growth of the three in our social networks over time, discovering that the two-against-one dynamic increases as relationships cement themselves. In summary, our novel system for predicting continuous tie strength values using general, platform-agnostic communication data establishes an innovative paradigm for studying the transitions and trends of interpersonal connections as they evolve in dynamic social networks. Moving forward, this paper will act as a foundation for our continued analysis into the evolution of relationships, and how they characterize past, present, and future interactions within social networks.

\section*{Methods}

\subsection*{Establishing Social Tie Rankings from Ego Network Surveys}

When implementing a tie strength measure, our foremost interest is choosing a system that allows for generalizability and customization, yet also produces a measure that is capable representing a nuanced spectrum of relationship strength. We additionally want to avoid a reliance on platform-specific attributes. While Marsden's work indicates that closeness is the best predictor of tie strength over other factors (specifically frequency of communication and duration of relationship)~\cite{marsden1984measuring}, we choose to avoid making any particular assumptions about predictor importance as well. Therefore, we introduce tie strength as an ranked list of individuals, where the ordering determines the depth of the relationship between a listed individual and the the participant that took the ego network survey.

In the ego network surveys the wording of the starting question that prompts a survey-taker to list individuals with whom they've communicated is void of any instructions on the ordering of said list. Thus, we cannot rely on the order of the raw list to be consistently indicative of a survey-taker's preference on any of the listed people. Given this, we assume there is none initially and instead craft our own using the follow-up survey questions and the staples of tie strength characterization as a guide. The answers to the follow-up questions are mostly selected from a set of answers that indicate a range of magnitude. For example, when asking after a survey-takers's perceived closeness with a listed individual they can choose "Especially close", "Close", "Less than close", or "Distant". Most questions follow this form, though there are a few questions like "How long in years have you known this person?" that have open inputs that can take any rational number. We utilized four inputs from our data as guided by the previously established definitions for tie strength: Closeness (how close the survey-taker is to one of the listed people), duration (how long the survey-taker has known the person), frequency (how often does the survey-taker communicate with the person), and similarity (subjectively, how similar does the survey-taker think themselves to be to the person). 

To determine an ordering from these mixed inputs without assuming the importance of any one input over another, we use a pairwise tournament selection process. Consider a ego network survey taken by one participant. Every individual listed by the participant is compared against all of the other listed individual on a question. An individual that has a greater value in the question than a counterpart is awarded a point. If two listed individuals have the same value, both are awarded a point. These points are aggregated across all the questions and then a ranked list is created by ordering everyone by their score in descending order. If there is a tie in aggregate score, the inputs with rational numbers are used to break the tie (e.g. for two tied individuals, the one who has been known for longer ultimately wins). After all tournaments are complete for that ego network survey, we are left with our top $k$ social ties ranking for the survey-taker, where the orderings indicate the importance of the listed individuals, as determined by the questions in the ego network survey. We repeat this process for all participants for all surveys throughout the study's timeline. Ultimately, the simplicity of this system ensures there are no assumptions made about the weighting of the questions. Additionally, this system is not dependent on a static set of features, since questions can be removed, replaced, or added and this won't change the architecture of how tie strength is generated in the end.

\subsection*{The Bow Tie Overlap Model}

Since overlapping (and non-overlapping) friend groups have shown to be a powerful tool in understanding tie strengths between people~\cite{bott2014family,mattie2018understanding}, implementing some measure of this kind of overlap is important to test its applicability within our datasets. Therefore, we introduce the weighted overlap metric from Mattie's work in bow tie frameworks~\cite{mattie2018understanding} given that (as mentioned earlier) this feature was highly informative for a couple of their tie strength prediction machine learning models. Weighted overlap is defined as below for two individuals $i$ and $j$:

\begin{equation}
    \widetilde{o}_{ij} = \frac{\sum_{k \in n_{ij}} (w_{ik} + w_{jk})}{s_i + s_j - 2 w_{ij}}
\end{equation}

Where $n_{ij}$ is the shared friends between $i$ and $j$. That is, the overlap in the $K=1$ neighbors of $i$ and $j$. We interpret the weights $w_{ij}$ here to be the total number of events between some $i$ and $j$ before the time of the survey being evaluated for reconstruction. And $s_i$ ($s_j$) is the total number of events generated by $i$ ($j$) before the time of the survey. Therefore, if all the individuals that have communicated with $i$ have also communicated with $j$ and vice versa, then $\widetilde{o}_{ij} = 1$. And so for some target individual $i$, we iteratively consider every communicated with individual as $j$ and sample each $\widetilde{o}_{ij}$ before the considered survey time. We then rank by value of $\widetilde{o}_{ij}$ to predict the ground truth survey ordering.





\subsection*{Machine Learning Models}

The primary models of this paper are our machine learning models. The machine learning models consider one target person at a time and make pairwise comparisons between the people with whom the target person has any communications history. Specifically, we consider the Ensemble model (which makes these pairwise comparisons with a random forest classifier), and the LSTM model (a two-channel long short-term memory recurrent neural network). For both models we use a method of ranking called Borda count~\cite{shah2017simple} for pairwise comparisons to generate the predicted ranked lists that we compare against the ground truth. This method is commonly viewed as ``an information-theoretically optimal procedure'' for recovering the top $k$ ranked items based on noisy comparisons that emphasizes simplicity, optimality, and robustness with regards to the underlying pairwise-comparison probability generation.

Given collection of $n$ people whom the target person has interacted indexed by the set $[n] \equiv \{1,...,n\}$, we create a matrix $M$ of dimensions $n \times n$ where $M_{ij}$ is the probability of $i$ having a greater social tie with the target person than $j$ as determined by the random forest or LSTM using $i$'s and $j$'s communication history with the target person up to the time of consideration. The diagonal of $M$, where $i = j$, is set to a probability of $\frac{1}{2}$. Now to find the Borda score, we must keep track of wins and losses in the pairwise tournament in $M$. To do this, we transform $M$ into $M'$ using Eq.~\ref{eq:m_prime}.

\begin{equation}\label{eq:m_prime}
    M'_{ij} = 
    \begin{cases}   
        1   & M_{ij} > \frac{1}{2} \cr
        0   & M_{ij} = \frac{1}{2} \cr
        -1  & M_{ij} < \frac{1}{2}
    \end{cases}
\end{equation}

The Borda count itself for $i \in [n]$, which is used to form the actual ranking, is calculated using Eq.~\ref{eq:borda_i}. 

\begin{equation}\label{eq:borda_i}
    B_i = \sum\limits_{j=1}^n M'_{ij}
\end{equation}

We then find $B_i$ for all $i \in [n]$ and then order by magnitude. This becomes the current predicted ranking that we compare against ground truth. Now, to generate the signals for Figure~\ref{fig:signals} and our network analysis we can convert the count to the winning percentage with Eq.~\ref{eq:borda_score}. In the equation, $w_i$, $l_i$, and $t_i$ are the number of head to head wins, losses, and ties for $i$. We generate the winning percentage incrementally across the entire NetSense study, and the resultant time series is then used as the dynamic edge weights between an ego and those they've communicated with.

\begin{equation}\label{eq:borda_score}
    \text{WinningPercentage}_{i} = \frac{B_i + (n-1)}{2(n-1)} = \frac{w_i + 0.5 \cdot t_i}{w_i + l_i + t_i}
\end{equation}

\subsubsection*{Ensemble Model}

The Ensemble model uses a random forest classifier to generate the pairwise comparison probabilities in $M$. The classifiers for the best performing Ensemble model used 100 weak classifiers. These classifiers are trained using a specific feature vector that is used to predict which individual will the target person have a greater social tie to. Consider the features for $i \in [n]$ (denoted as $f_i$) as the four baseline class features computed for just calls and just texts. These features are frequency, recency, duration, and volume as described in Models section. Integrating the Bow Tie Overlap attribute significantly brings down overall performance, and so was excluded in the final Ensemble model. We take the difference of these two feature vectors as the given feature vector for the classifiers, defined as $\text{DifferenceFeatureVector}(x,y) = f_x - f_y$ given $x,y \in [n]$.

\subsubsection*{LSTM Model}

For our LSTM models, $f_i$ is a two-channel time series. The two channels are the histories of calls and texts, both binned into 21 days intervals. The feature vector used by the LSTM is the time series for the two individuals being compared stacked on each other, resulting in a four channel time series that spans through time at which the social tie is being evaluated to the first interaction between the target person and either person in the comparison.

\bibliographystyle{unsrt}  
\bibliography{references}

\begin{thebibliography}{10}

\bibitem{borgatti2009network}
Stephen~P Borgatti, Ajay Mehra, Daniel~J Brass, and Giuseppe Labianca.
\newblock Network analysis in the social sciences.
\newblock {\em science}, 323(5916):892--895, 2009.

\bibitem{kitts2020rethinking}
James~A Kitts, Eric Quintane, and ESMT Berlin.
\newblock Rethinking social networks in the era of computational social
  science, 2020.

\bibitem{rivera2010dynamics}
Mark~T Rivera, Sara~B Soderstrom, and Brian Uzzi.
\newblock Dynamics of dyads in social networks: Assortative, relational, and
  proximity mechanisms.
\newblock {\em annual Review of Sociology}, 36:91--115, 2010.

\bibitem{granovetter1977strength}
Mark~S Granovetter.
\newblock The strength of weak ties.
\newblock In {\em Social networks}, pages 347--367. Elsevier, 1977.

\bibitem{krackhardt2003strength}
David Krackhardt, N~Nohria, and B~Eccles.
\newblock The strength of strong ties.
\newblock {\em Networks in the knowledge economy}, 82, 2003.

\bibitem{marsden1984measuring}
Peter~V Marsden and Karen~E Campbell.
\newblock Measuring tie strength.
\newblock {\em Social forces}, 63(2):482--501, 1984.

\bibitem{policarpo2015friend}
Ver{\'o}nica Policarpo.
\newblock What is a friend? an exploratory typology of the meanings of
  friendship.
\newblock {\em Social Sciences}, 4(1):171--191, 2015.

\bibitem{onnela2007structure}
J-P Onnela, Jari Saram{\"a}ki, Jorkki Hyv{\"o}nen, Gy{\"o}rgy Szab{\'o}, David
  Lazer, Kimmo Kaski, J{\'a}nos Kert{\'e}sz, and A-L Barab{\'a}si.
\newblock Structure and tie strengths in mobile communication networks.
\newblock {\em Proceedings of the national academy of sciences},
  104(18):7332--7336, 2007.

\bibitem{jones2013inferring}
Jason~J Jones, Jaime~E Settle, Robert~M Bond, Christopher~J Fariss, Cameron
  Marlow, and James~H Fowler.
\newblock Inferring tie strength from online directed behavior.
\newblock {\em PloS one}, 8(1), 2013.

\bibitem{gilbert2009predicting}
Eric Gilbert and Karrie Karahalios.
\newblock Predicting tie strength with social media.
\newblock In {\em Proceedings of the SIGCHI conference on human factors in
  computing systems}, pages 211--220, 2009.

\bibitem{conti2011a_model}
M.~{Conti}, A.~{Passarella}, and F.~{Pezzoni}.
\newblock A model for the generation of social network graphs.
\newblock In {\em 2011 IEEE International Symposium on a World of Wireless,
  Mobile and Multimedia Networks}, pages 1--6, 2011.

\bibitem{wiese2015you}
Jason Wiese, Jun-Ki Min, Jason~I Hong, and John Zimmerman.
\newblock "you never call, you never write" call and sms logs do not always
  indicate tie strength.
\newblock In {\em Proceedings of the 18th ACM conference on computer supported
  cooperative work \& social computing}, pages 765--774, 2015.

\bibitem{mattie2018understanding}
Heather Mattie, Kenth Eng{\o}-Monsen, Rich Ling, and Jukka-Pekka Onnela.
\newblock Understanding tie strength in social networks using a local “bow
  tie” framework.
\newblock {\em Scientific reports}, 8(1):1--9, 2018.

\bibitem{purta2016experiences}
Rachael Purta, Stephen Mattingly, Lixing Song, Omar Lizardo, David Hachen,
  Christian Poellabauer, and Aaron Striegel.
\newblock Experiences measuring sleep and physical activity patterns across a
  large college cohort with fitbits.
\newblock In {\em Proceedings of the 2016 ACM international symposium on
  wearable computers}, pages 28--35, 2016.

\bibitem{blair2015cell}
Bethany~L Blair, Anne~C Fletcher, and Erin~R Gaskin.
\newblock Cell phone decision making: Adolescents’ perceptions of how and why
  they make the choice to text or call.
\newblock {\em Youth \& Society}, 47(3):395--411, 2015.

\bibitem{homans_1950}
George Homans.
\newblock {\em The Human Group}, page 133.
\newblock Harcourt, Brace \& World, 1950.

\bibitem{bott2014family}
Elizabeth Bott and Elizabeth~Bott Spillius.
\newblock {\em Family and social network: Roles, norms and external
  relationships in ordinary urban families}.
\newblock Routledge, 2014.

\bibitem{breiman2001random}
Leo Breiman.
\newblock Random forests.
\newblock {\em Machine learning}, 45(1):5--32, 2001.

\bibitem{hochreiter1997long}
Sepp Hochreiter and J{\"u}rgen Schmidhuber.
\newblock Long short-term memory.
\newblock {\em Neural computation}, 9(8):1735--1780, 1997.

\bibitem{webber2010similarity}
William Webber, Alistair Moffat, and Justin Zobel.
\newblock A similarity measure for indefinite rankings.
\newblock {\em ACM Transactions on Information Systems (TOIS)}, 28(4):1--38,
  2010.

\bibitem{seshadhri2012community}
Comandur Seshadhri, Tamara~G Kolda, and Ali Pinar.
\newblock Community structure and scale-free collections of
  erd{\H{o}}s-r{\'e}nyi graphs.
\newblock {\em Physical Review E}, 85(5):056109, 2012.

\bibitem{chen2009extending}
Yi-Fan Chen and James~E Katz.
\newblock Extending family to school life: College students’ use of the
  mobile phone.
\newblock {\em International Journal of Human-Computer Studies},
  67(2):179--191, 2009.

\bibitem{saramaki2014persistence}
Jari Saram{\"a}ki, Elizabeth~A Leicht, Eduardo L{\'o}pez, Sam~GB Roberts, Felix
  Reed-Tsochas, and Robin~IM Dunbar.
\newblock Persistence of social signatures in human communication.
\newblock {\em Proceedings of the National Academy of Sciences},
  111(3):942--947, 2014.

\bibitem{roberts2011costs}
Sam~GB Roberts and Robin~IM Dunbar.
\newblock The costs of family and friends: an 18-month longitudinal study of
  relationship maintenance and decay.
\newblock {\em Evolution and Human Behavior}, 32(3):186--197, 2011.

\bibitem{caplow1968two}
Theodore Caplow.
\newblock {\em Two against one: Coalitions in triads.}
\newblock Prentice-Hall, 1968.

\bibitem{shah2017simple}
Nihar~B Shah and Martin~J Wainwright.
\newblock Simple, robust and optimal ranking from pairwise comparisons.
\newblock {\em The Journal of Machine Learning Research}, 18(1):7246--7283,
  2017.

\end{thebibliography}

\section*{Acknowledgements}

This work was sponsored in part by DARPA under contract W911NF-17-C-0099, the Army Research Office (ARO) under contract W911NF-17-C-0099, and the Office of Naval Research (ONR) under grant N00014-15-1-2640. The views and conclusions contained in this document are those of the authors and should not be interpreted as representing the official policies either expressed or implied of the U.S. Government.

\section*{Author contributions statement}

B.K.S and J.F. conceived the study. J.F. and R.D. formalized and implemented all models. B.K.S., J.F., and R.D. analyzed model performance. J.F. designed and implemented all evolving tie strength analytical tests and evaluated the results. B.K.S., J.F., and R.D. wrote the first draft of the manuscript. All authors read and approved the paper.

\newpage

\renewcommand{\thetable}{S\arabic{table}}
\setcounter{table}{0}   

\renewcommand{\thefigure}{S\arabic{figure}}
\setcounter{figure}{0}

\end{document}